\documentclass[aps,prl,twocolumn,groupedaddress,showpacs]{revtex4}

\usepackage{dcolumn}
\usepackage{graphicx,amssymb,amsmath}
\usepackage{bm}
\usepackage{natbib}
\usepackage{epstopdf}
\begin{document}

\title{An Anderson Transition of the Plasma Oscillations of 1D Disordered Wigner Lattices}

\author{Shimul Akhanjee}
\email[]{shimul@physics.ucla.edu}
\author{Joseph Rudnick}
\email[]{jrudnick@physics.ucla.edu}
\affiliation{Department of Physics, UCLA, Box 951547 Los Angeles, CA 90095-1547}


\date{\today}

\begin{abstract}
We report the existence of a localization-delocalization transition in the classical plasma modes of a one dimensional Wigner Crystal with a white noise potential environment at $T=0$. Finite size scaling analysis reveals a divergence of the localization length at a critical eigenfrequency. Further scaling analysis indicates power law behavior of the critical frequency in terms of the relative interaction  strength of the charges. A heuristic argument for this scaling behavior is consistent with the numerical results. Additionally, we explore a particular realization of random-bond disorder in a one dimensional Wigner lattice in which all of the collective modes are observed to be localized.
\end{abstract}


\maketitle

Recently, there has been considerable interest in the localization phenomena of disordered, one dimensional(1D), electronic tight-binding models. Moreover, it has been established that an Anderson transition can occur in lower dimensional systems containing nearest-neighbor interactions and random on site energies exhibiting long range correlations in its power spectrum~ \cite{izrailev:1dmob1}. Relatively little is firmly understood about the classical oscillator models with long range interactions and disorder. Wigner Crystal(WC) \cite{wigner:1dmob1} phases are one such example, having collective excitations that posess a different character from the weakly interacting regime, especially in the presence of disorder. Various authors have studied plasmon localization in this context. However those models do not consider the true long range nature of the coloumb potential as discussed in this article \cite{fogler:1dmob1}. In this extreme limit of electron-electron interactions, the dominant behavior of the collective modes can be treated classically due to the negligible overlap of the electronic wavefunctions at lower densities~ \cite{giamarchi:1dmob1}.

In this paper we report on the existence of a localization-delocalization transition in the plasma oscillations of a disordered 1D WC at zero temperature. We provide strong evidence for the existence of such a transition based on numerical studies of both the localization length $\xi$ and the inverse participation ratio $P$. In particular we show  that the transition occurs in a 1D system with \emph{uncorrelated} disorder in contrast to the mobility edge observed by Izrailev et al~ \cite{izrailev:1dmob1}. Data collapse at various system sizes confirms both scaling at the transition and a critical exponent $\nu$ associated with the divergence of the localization length.  Additionally, we recover a scaling exponent of the critical eigenfrequency in terms of the relative  interaction strength of the charges. We also explore a disordered model with randomness in the effective bonds between the charges. This can be viewed as an extension of Dyson's disordered chain problem~ \cite{dyson:1dmob1}. In the case of this system no Anderson-like transition exists. Rather, all plasma oscillation eigenstates are localized. In this regard, the system mirrors the behavior of the one-dimensional system with short-range disorder. Unlike that system,  the localization length is not well described by the Thouless formula which follows from the mathematical simplicity of a dynamical matrix having tri-diagonal structure~ \cite{thouless:1dmob1}.

The first system under investigation, which we term Model A, consists of like charges placed in an external random potential with the charge neutrality condition enforced by smearing a positive background over the chain. The Hamiltonian for Model A is given by:
\begin{equation}
H = \sum\limits_{i = 1}^L {\frac{{p_i^2 }}{{2m_e }}}  + \frac{J}{2}\sum\limits_{i \ne j} {\frac{{Q_i Q_j }}{{\left| {x_i  - x_j } \right|}}} + \sum\limits_i^L {Q_i V(x_i)}
\label{eq:ham}
\end{equation}
We assume the low density limit in which Coloumb interactions dominate the kinetic energy term, resulting in a Wigner solidification of the particle array. $V(x)$ is a white noise potential with the following Fourier decomposition, with $N = L/4$ Fourier components.
\begin{equation}
V(x) = A\sum\limits_n^N {a_n } \cos (2\pi nx) + b_n \sin (2\pi nx)
\label{eq:fourier}
\end{equation}
The parameters $J$ and $A$ in Eqs. (\ref{eq:ham}) and (\ref{eq:fourier}) are dimensionless coupling constants, which we utilize to define the dimensionless interaction strength $\kappa \equiv {J/A} $.
 The random variables $a_n$ and $b_n$ are chosen from different Gaussian distributions, having an independence of $n$, yielding a white noise power spectrum with a mean of $\mu=0$ and a variance of $\sigma=1$. In Model B the external potential, $V(x)=0$ and the charges $\{Q_i\}$ are random variables with a distribution function $P[Q]$, given by:
\begin{equation}
P[Q] = \left\{ \begin{array}{ll}
 \frac{1}{{W_2  - W_1 }} & W_1  < Q < W_2  \\
 0 & {\rm otherwise} \\
 \end{array} \right.
 \label{eq:Qdist}
\end{equation}

The formalism used for computing the eigenfunctions of the collective modes involves the construction of the dynamical matrix $\mathbf{D(R)}$. For a finite sized chain of length $L$, $\mathbf{D(R)}$ is an $L \times L$ symmetric matrix. In the harmonic approximation the full electrostatic potential is expanded in a Taylor series with the terms proportional to the second derivatives retained. The dynamical matrix has the structure
\begin{equation}
\mathbf{D(R - R')} = \delta _{\mathbf{R,R'}} \sum\limits_{\mathbf{R''}} {\left. {\frac{{\partial ^2 \phi (x)}}{{\partial x^2 }}} \right|} _{x = \mathbf{R - R''}}
  - \left. {\frac{{\partial ^2 \phi (x)}}{{\partial x^2 }}} \right|_{x = \mathbf{R - R'}}
  \label{eq:D(R)}
\end{equation}
where $\phi (x)$ is defined as the electrostatic potential between two charges in a periodic image and the R's are the equilibrium positions of the charges to be determined by a numerical minimization. The dynamical matrix enters into the the eigenvalue equation for plasma eigenmodes as follows  \cite{AM}:
\begin{equation}
m_e \omega ^2 u(\mathbf{R}) + \sum\limits_{\mathbf{R'}} {\mathbf{D(\mathbf{R - R'})}} u(\mathbf{R'}) = 0
\label{eq:phieq}
\end{equation}
where the functions $u(\mathbf{R})$ are the lattice displacements from equilibrium. Periodic boundary conditions were employed to eliminate surface effects, and lengths are scaled so that the size of the region occupied by the charges is unity. We make use of a closed form expression for the electrostatic potential, $\phi(x)$, that possesses the correct periodicity, short distance behavior and that properly reproduces the standard Ewald potential result to high degree of accuracy. The explicit form that we utilized is
\begin{equation}
\phi (x) = \pi |\csc (\pi x)|
\label{eq:coulint}
\end{equation}

After assigning charge values to the particles in an ordered array,  the system must relax so that the total force on each particle is zero. We employed the Newton-Raphson method to numerically determine the particle coordinates at equilibrium.

Given the relaxed array of charges, we are in a position to calculate the dynamical matrix $\mathbf{D(R)}$ via (\ref{eq:D(R)}) and the solutions of the eigenfunction equation (\ref{eq:phieq}).
The eigenfunction width $w$ is a direct measure of the localization length $\xi$, taken as the  mean squared deviation of the mimimum distance between lattice sites.
A natural choice is:

\begin{eqnarray}
 w_i & = & \frac{1}{2}\sum_{k,l=1}^L u_i(k)^2 u_i(l)^2(k-l)^2  \nonumber \\ & = &  \sum\limits_{k = 1}^L {u_i (k)^2 k^2 }  - \left( {\sum\limits_{k = 1}^L {u_i (k)^2 k} } \right)^2 \nonumber   \\
  &= & \left\langle {k^2 } \right\rangle _i  - \left( {\left\langle k \right\rangle _i } \right)^2  = \left\langle {\left( {k - \left\langle k \right\rangle _i } \right)^2 } \right\rangle _i
\label{eq:width}
\end{eqnarray}
where $u_i (k)$ is the amplitude of the ith eigenfunction at the site $k$. The one difficulty with this definition of the width is that a function localized to the boundaries of the interval would be described as extended. To correctly take into account the periodicity of the system and to avoid the above pitfall, we replace $(k-l)^2$ in the first line of (\ref{eq:width}) by $\sin^2 \pi(k-l)/\pi^2$.

Another important measure of spatial localization is  the inverse participation ratio (IPR) $P$ \cite{wegner:1dmob1}, defined as~

\begin{eqnarray}
P_i  = L{\sum\limits_{j = 1}^L {\left| {u_i (j)} \right|^4 } }/{( {\sum\limits_{j = 1}^L {\left| {u_i (j)} \right|^2 } } )^2 } = L{\sum\limits_{j = 1}^L {\left| {u_i (j)} \right|^4 } }
\label{eq:ipr}
\end{eqnarray}

The quantity $P$ is a useful measure of which fraction of the oscillators
are localized. As the thermodynamic limit is taken, $P$ approaches zero for extended modes, while for localized modes the IPR can be as high as $P=L$.
For both models discussed in this article, one must account for the numerical and statistical uncertainties on finite length chains that scale like $\propto {1 \mathord{\left/{\vphantom {1 {\sqrt L }}} \right.\kern-\nulldelimiterspace} {\sqrt L }}
$. Therefore, $w$ and $P$ were ensemble averaged over 50 realizations of disorder to smooth out these fluctuations.

\begin{figure}
\centerline{\includegraphics[height=2.3in]{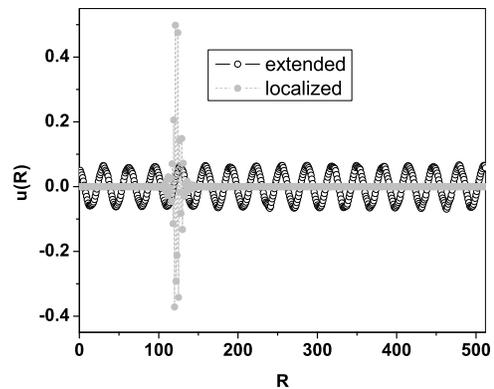}}
\caption{Typical extended and localized eigenfunctions(Model A, L=512)}
\label{fig:typical}
\end{figure}

Typical eigenfunctions in the extended and localized regimes of Model A are shown in Fig. \ref{fig:typical}. Due to the limitation of performing diagonalizations of finite sized matrices, one can interpret an extended eigenfunction as having a width value, $w_i =\xi = L$.
The existence of criticality is indicated by a common crossing point for the values of $w$ and $P$ that are normalized at various system sizes.
A genuine Anderson transition is usually accompanied by a diverging localization length at a particular eigenvalue~ \cite{anderson:1dmob1} and a similiar abrupt shift in the IPR at this same eigenvalue. For electronic tight binding systems, the critical eigenvalue that separates conducting and insulating phases  is known as the mobility edge~ \cite{mott:1dmob1}. Likewise, in Model A the plasma oscillations display two frequency regimes, one corresponding to localized and the other to extended lattice waves analogous to the mobility edge. However the DC conductivity for charge transport is  equal to zero as Wigner crystals are known to be Mott Insulators~ \cite{wigner:1dmob1}. Furthermore, the interaction matrix elements of classical oscillator systems are not constrained to even integer sites as usually implemented in tight binding systems with random hopping in order to preserve the particle-hole symmetry of the Hamiltonian.
Therefore along with our required structural relaxation procedure, the  Models A and B fundamentally differ from those investigated by Bhatt and Zhou~ \cite{bhatt:1dmob1}.

\begin{figure}
\centerline{\includegraphics[height=4in]{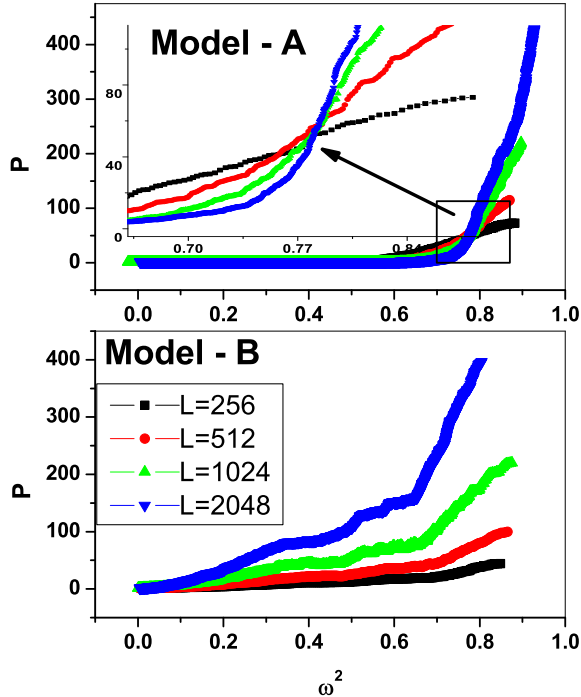}}
\caption{Inverse participation ratio vs. eigenvalues normalized at different system sizes showing transition point in Model A.}
\label{fig:ipdata}
\end{figure}

The IPR data is presented in Fig. \ref{fig:ipdata}. For all cases in Model A, the interaction strength parameter, $\kappa$ is set equal to $0.1$.
Note that for a range of low frequencies, the localization lengths saturate at $\xi=L$, while at higher frequencies, $\xi \to 0$. In fact, Fig. \ref{fig:ipdata} Model A appears to be consistent with a transition between frequency regimes in which the lattice excitations shift between the general forms illustrated in Fig. \ref{fig:typical}.
Additionally, Fig. \ref{fig:ipdata} exhibits a common crossing point in Model A that is consistent with asymptotically diverging IPR values above the crossing point as a function of L, signifying localization.

\begin{figure}
\centerline{\includegraphics[height=2.5in]{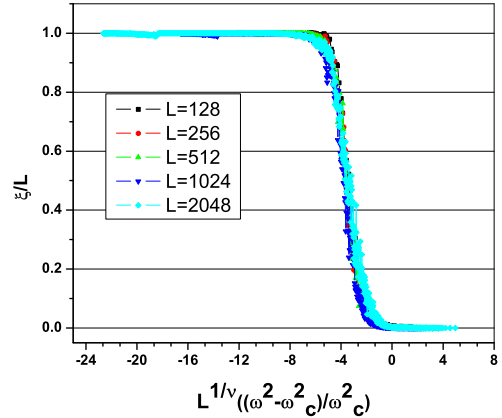}}
\caption{Normalized localization length vs. scaled eigenvalues for different system sizes, showing data collapse at $1 / \nu \approx 0.45 \pm 0.05$ (Model A)}
\label{fig:scaling}
\end{figure}

In another test for a transition, we attempt the application of finite size scaling analysis to the dependence of the localization length of the lattice modes of the Model A system  \cite{fishergreen,privman}. The results of this analysis are shown  in Fig. \ref{fig:scaling}. We test for a scaling form for the localization length having the following dependence on key quantities
\begin{equation}
\xi /L \propto F_a (L^{1/\nu } (\omega ^2  - \omega _c^2 )/\omega _c^2 )
\label{eq:scalingform}
\end{equation}
The function $F_a$ is a universal function present in phase transitions belonging to the same universality class, the precise nature of which has not yet been determined. One sees a clear collapse of the data at $1 / \nu \approx 0.45 \pm 0.05$. We performed a statistical analysis of the curves at different system sizes to provide a quantitative measure of the "goodness" of the collapse. Our results yield a chi squared p value of
$p =0.37$, indicating no statistically significant differences between the collapsed curves.
For fixed $\kappa$, the critical frequency $\omega_c$ is asymptotically independent of system size in the limit of a large system. The dependence of the square of the critical frequency $\omega _c^2$ on the relative interaction strength was examined between the values of $ \kappa = 0.075 $ and $ \kappa = 0.1 $, for the values of  $\omega _c^2$ within the center of the band. In this regime the Coulomb force is close to the random force in strength. We performed a scaling analysis in this non-perturbative regime, assuming a scaling form of  $\omega _c^2  = A\kappa ^\delta $. We have determined $\delta \approx 0.55 \pm 0.008 $ from a linear fit with an R squared value of 0.999.

We can attempt to qualitatively understand this exponent  by considering the Hamiltonian containing the random potential and the unscreened coloumb interactions. We speculate that the localization-delocalization can be attributed to the threshold at which the range of the Coloumb force is comparable to the length scale of the net disorder force. We define this length scale $\overline r _0$. The explicit scale dependence of the Coloumb force and the pinning forces are given as
\begin{equation}
F_{coulomb}  \propto \frac{\kappa }{{\overline r _0^2 }}
\label{eq:coulkappa}
\end{equation}
\begin{equation}
F_{ran}  \propto \overline r _0^{1/2}
\end{equation}
Clearly, the net random force should follow a random walk that scales with the number of pinning  centers or in terms of the characteristic length scale, the length between the particles. It follows that at the threshold,
\begin{equation}
\overline r _0^{}  \propto \kappa ^{2/5}
\end{equation}

The critical eigenvalues will then scale as,

\begin{equation}
\omega _c^2  \sim \frac{\kappa }{{\overline r _0^{} }} \sim \kappa ^{3/5}
\end{equation}

We find it rather interesting that this value of $\delta =0.6$ compares well to our numerically determined value of $\delta \approx 0.55 \pm 0.008$. It is possible that this scale dependent behavior can be connected to the collective pinning of elastic manifolds at the Larkin scale \cite{larkin:1dmob1}.
The low frequency dynamics of pinned elastic systems has been studied by Fogler, where it is argued that the localization length for plasmons is related to the Larkin length
defined as the size $R_c$ of domains individually pinned by spatially random potential wells \cite{fogler:1dmob1}.

On the other hand, model B appears to admit of no Anderson transition as shown in Figs. \ref{fig:ipdata}. All oscillator systems contain the extended eigenmode corresponding to the $\omega^2 = 0$ eigenvalue. For Model B as the thermodynamic limit is approached, $\xi$ asymptotically approaches this one extended mode while the rest of the modes are localized. Additionally, no crossing point exists in the IPR data, confirming an absence of criticality. We have checked this result for various distribution widths and mean values defined in (\ref{eq:Qdist}), and no criticality has been observed. For both models one can attempt to understand the precise behavior of the eigenwidths in terms of the Lyapunov exponent $\gamma = 1/ 2\xi$ derived from the residue of the single  particle Green's function \cite{thouless:1dmob1}.
\begin{equation}
\gamma _\beta   = \int {g( \omega^2 )\ln \left| {\omega^2_\beta   - \omega^2} \right|} d\omega^2 - \frac{{\ln \left| {\mathop{\rm Cof}(\omega^2\mathbb{I} - D)_{L1} } \right|}}
{{L - 1}}
\label{eq:gamma}
\end{equation}

In Eq. (\ref{eq:gamma}), $g( \omega^2 )$ is the density of normal modes and Cof is the matrix cofactor of the inverse resolvent. For systems with nearest-neighbor interactions the second term in Eq. (\ref{eq:gamma}) does not depend on the eigenvalues and averages to a constant yielding the Thouless formula~ \cite{thouless:1dmob1}. However, for systems with long-ranged interactions this term exhibits a nontrivial eigenvalue dependence. An Anderson transition would correspond to $\gamma(\omega^2_c)= 0$. Ultimately, the existence of unscreened $1/R$  interactions is simply not the sole factor responsible for the transition. Rather it is the complex and  subtle interplay of disorder and interactions.

We conclude by noting that the localization properties of lattice waves has important consequences with regard to the energy and heat transport of real physical systems. Recent articles have discussed a new class of materials that contain structurally ordered charge configurations such as the copper-oxide chains of compounds like Na$_{1+x}$CuO$_2$~ \cite{mayr:1dmob1}. These materials contain magnetic and thermodynamic properties that are consistent with some type of 1D Wigner lattice  formation. A possible thermodynamic signature of an Anderson transition observed in Model A would be present in the heat transport, given that a major source of the carriers of energy would arise from the collective excitations of the electronic lattice. \cite{alex}

Finally, we note that eigenfunction localization has been observed in random banded matrices with power law modulation by Mirlin et. al.  \cite{mirlin} and by Levitov  \cite{levitov} in one-dimensional tight-binding Hamiltonians with long-range hopping.The Hamiltonian governing the dynamics of plasmons in our model shares important features with those systems in that in all three cases the issue is the spatial structure of the eigenfunctions of matrices that incorporate randomness and that contain entries that decay as a power law in the difference between indices. However, key power laws are different, in that we observe an effective mobility edge in eigenvectors of an operator in which off-diagonal terms decay effectively as $|i-j|^{-3}$, rather than as the inverse first power, as in the case of  \cite{levitov,mirlin}.

\begin{acknowledgments}
The authors would like to acknowledge the many useful conversations with S.E. Brown, H.W. Jiang, and
S. Chakravarty. We also give our thanks to M. Fogler for correspondence and useful suggestions.

\end{acknowledgments}

\end{document}